\documentclass[twocolumn,showpacs,amsmath,aps,pra,amssymb,nofootinbib,superscriptaddress]{revtex4-1} 
\usepackage[T1]{fontenc}
\usepackage[latin9]{inputenc}
\usepackage{times}
\usepackage{color} 
\usepackage{xspace}
\usepackage{amssymb,amsmath}
\usepackage{amsbsy}
\usepackage{graphicx}
\usepackage{bm}
\usepackage{epstopdf}
\usepackage{float}
\usepackage{adjustbox}

\usepackage[unicode,breaklinks]{hyperref}
\hypersetup{
    unicode=true,
    a4paper=true,
    plainpages=false, 
    colorlinks=true,
    linkcolor=blue,
    citecolor=blue,
    filecolor=black,
    urlcolor=blue
}
\newcommand{\od}[1]{\ensuremath{{O\negthickspace\left({{#1}}\right)}}\xspace}
\newcommand{\kt}{k_B T}
\newcommand{\bk}{\mathbf{k}}

\newcommand{\br}{\mathbf{r}}
\newcommand{\bx}{\mathbf{x}}
\newcommand{\taut}{\tilde{\tau}_\sigma}
\newcommand{\brho}{\bm{\rho}}
\newcommand{\efree}{\epsilon_\bk}

\newcommand{\Vdir}{\nabla_\bk \tilde{V}(\mathbf0)}

\begin{document}
%\title{Anisotropic  number fluctuations of a dipolar Bose-Einstein condensate} 
\title{Number Fluctuations of a Dipolar Condensate:\\ Anisotropy and Slow Approach to the Thermodynamic Regime}
\author{D.~Baillie}
\affiliation{Jack Dodd Centre for Quantum Technology, Department of Physics, University of Otago, Dunedin 9016, New Zealand}
\author{R.~N.~Bisset}
\affiliation{Center for Nonlinear Studies and Theoretical Division, Los Alamos National Laboratory, Los Alamos, New Mexico 87545, USA}
\author{C.~Ticknor}
\affiliation{Center for Nonlinear Studies and Theoretical Division, Los Alamos National Laboratory, Los Alamos, New Mexico 87545, USA}
\author{P.~B.~Blakie}  
\affiliation{Jack Dodd Centre for Quantum Technology, Department of Physics, University of Otago, Dunedin 9016, New Zealand}

\begin{abstract}
    We present a   theory for the number fluctuations of a quasi-two-dimensional (quasi-2D) dipolar Bose-Einstein condensate measured with finite resolution cells. We show that when the dipoles are tilted to have a component parallel to the plane of the trap,  the  number fluctuations become anisotropic, i.e.~depend on the in-plane orientation of the measurement cell. We develop analytic results for the quantum and thermal fluctuations  applicable to the cell sizes  accessible in experiments. We show that as  cell size is increased the  thermodynamic  fluctuation result is approached much more slowly  than in condensates with short range interactions, so experiments would not require high numerical aperture imaging to observe the predicted effect.
  \end{abstract}

\maketitle

Many important aspects of a quantum system are revealed by studying its fluctuations. 
For instance, in quantum optics  intensity fluctuations   distinguish a laser from chaotic light \cite{hbt}, and reveal nonclassical effects such as photon antibunching in resonance florescence \cite{Carmichael1976a}.
With the rapid progress made in the experimental study of quantum gases, the measurement of fluctuations has become of increasing interest  (e.g.~see \cite{Altman2004a,Folling2005a,Greiner2005a,Jeltes2007a,Hofferberth2008a,Muller2010a,Sanner2011a,Cheneau2012a,Hung2011,Jacqmin2011a,Armijo2012a,Blumkin2013a}). Particularly for the insight it  provides into manybody physics, including nonequilibrium regimes where it  is a critical tool for studying quenches \cite{Cheneau2012a,Hung2013}.

Here we study the fluctuations of a dipolar BEC, a system that promises a diverse range of new physics   (e.g.~\cite{Baranov2008,Lahaye_RepProgPhys_2009}) and has seen remarkable experimental progress with the production of chromium \cite{Griesmaier2005a,Beaufils2008a}, dysprosium \cite{Mingwu2011a} and erbium \cite{Aikawa2012a} condensates. The key new feature of these systems is that the atoms interact via a dipole-dipole interaction (DDI) that is both long ranged and anisotropic. The anisotropy manifests itself in the condensate density distribution (magnetostriction) \cite{Stuhler2005a,Lahaye2008a} and the excitation spectrum  \cite{Bismut2012a}, and thus in the system coherence functions  \cite{Ticknor2012b}, superfluid properties  \cite{Ticknor2011a}, and critical temperature \cite{Glaum2007}. Additionally, the DDI is partially attractive and   produces rotonlike excitations in confined geometries \cite{Santos2003a,Ronen2007a,Macia2012b} and  enhances  density fluctuations \cite{Klawunn2011,Boudjemaa2013a,Bisset2013a,Blakie2013a}.

The atom number fluctuation measurements are performed in finite-sized cells, with minimum cell dimension being limited by the imaging system resolution (typically  much larger than the BEC healing length). The crucial role that the cell plays in the measurement has been the subject of recent attention \cite{Hung2011a,Jacqmin2011a,Armijo2012a}. Tailoring the size and shape of the cell (e.g.~by amalgamating the signal collected on detector pixels \cite{Armijo2012a}) can be used to identify localised roton modes \cite{Bisset2013a} and at sufficiently low temperatures (and for cells smaller than the thermal wavelength)  the measurements will be dominated by quantum fluctuations of collective modes \cite{Armijo2012a},   predicted to exhibit nonextensive scaling with cell size \cite{Astrakharchik2007a,Klawunn2011}.

\begin{figure}[ht!] 
   \centering
  \vspace*{-0.25cm}
  \includegraphics[width=2.3in]{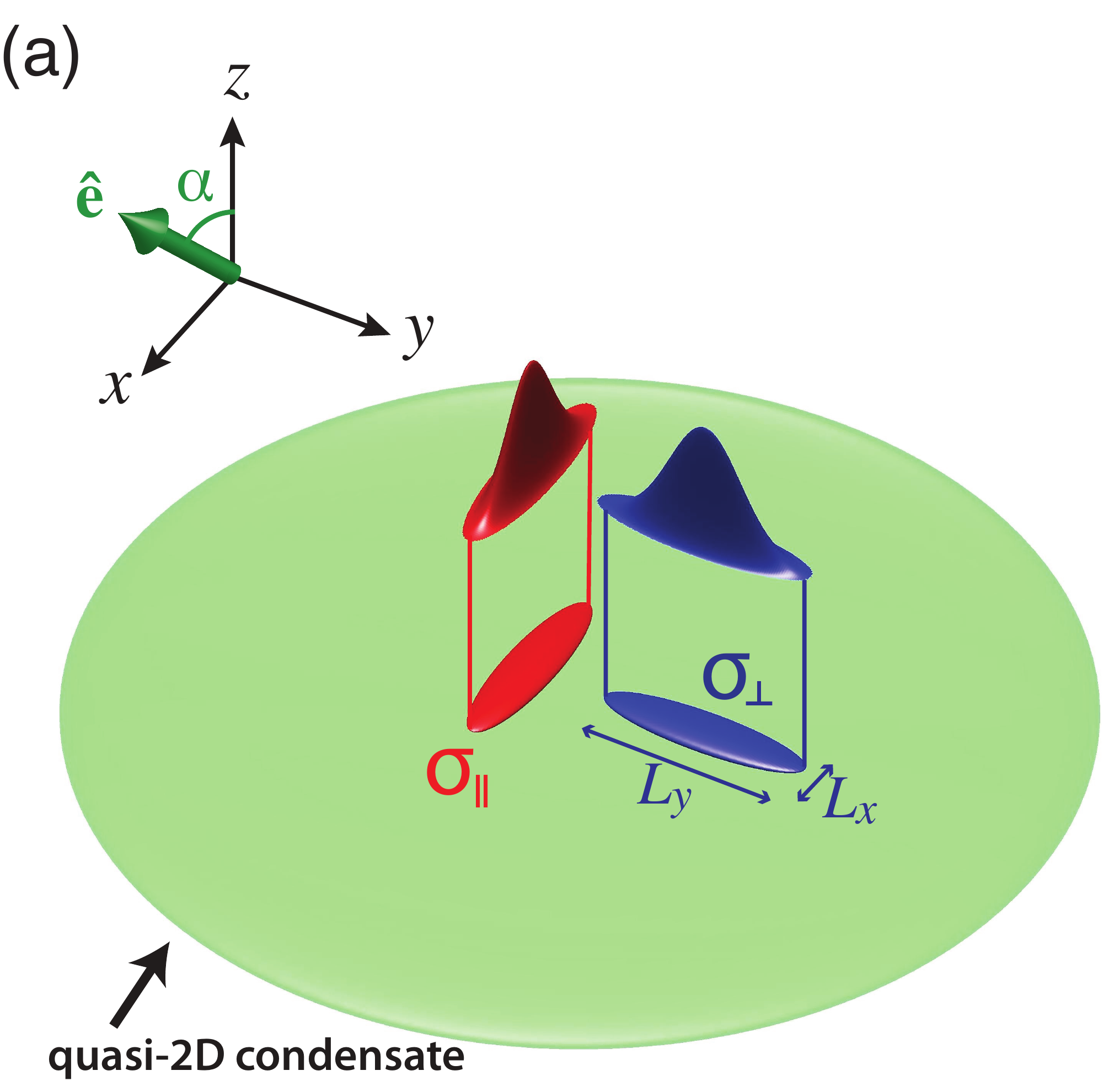}       \vspace*{-0.0cm}\\
\includegraphics[width=3.0in]{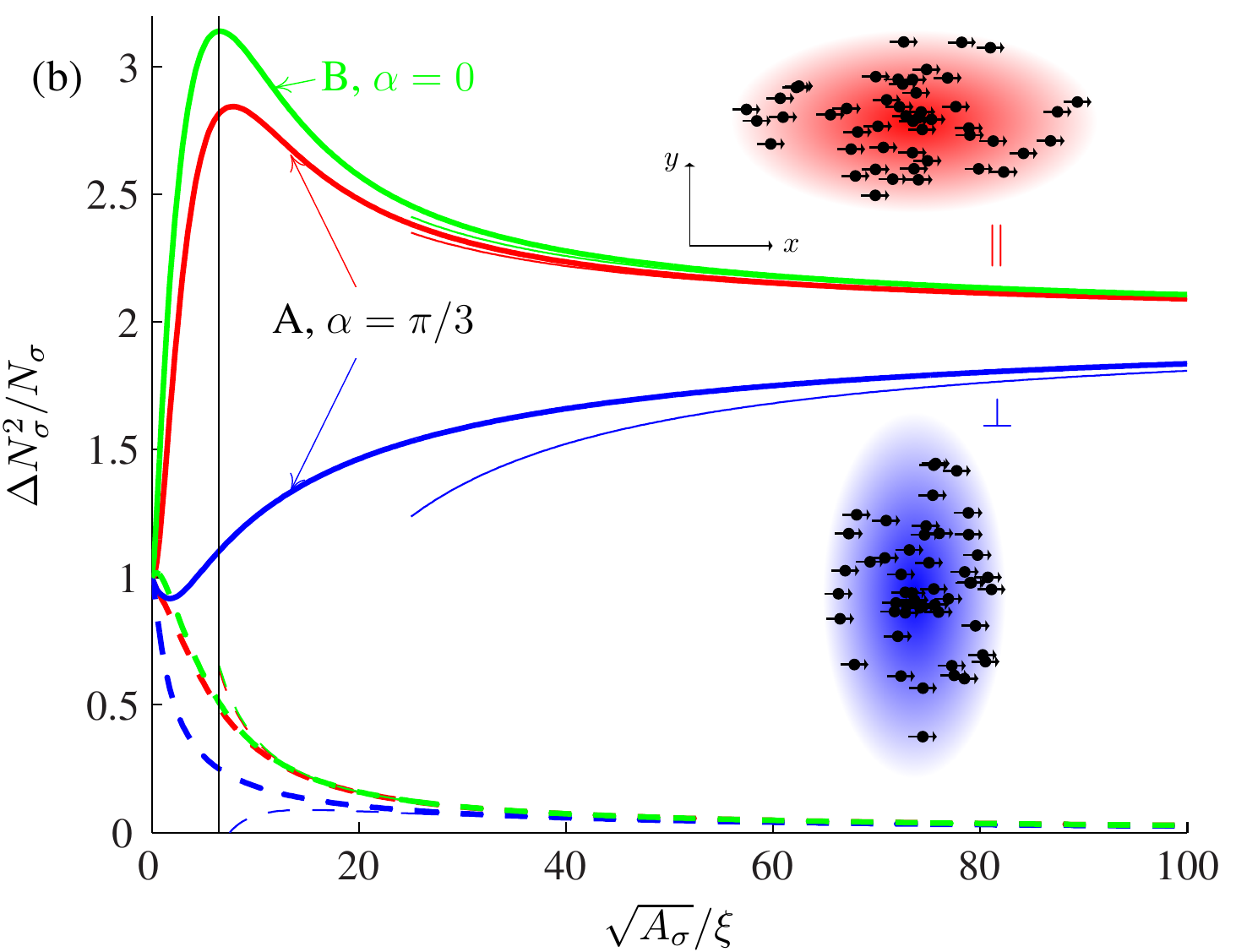}
  \vspace*{-0.25cm}
   \caption{(color online)  (a)  Schematic: a Gaussian cell $\sigma_\perp$ of size $L_x$-by-$L_y$ is used to measure atom number in a   quasi-2D dipolar BEC. Anisotropy in number fluctuations is revealed by a second cell $\sigma_\|$  identical to $\sigma_\perp$, but rotated by 90$^\circ$ (offset in location for clarity).   The   dipoles are polarised along $\hat{\mathbf{e}}$, which is tilted in the $xz$-plane at an angle of $\alpha$ to the $z$ axis.  
   (b) Fluctuations as a function of cell area for  dipolar BECs with  $\alpha=\pi/3$, aspect ratio $\lambda=L_y/L_x=0.15$ (red, $\|$) and $\lambda=1/0.15$ (blue, $\perp$) and $\alpha=0$ (green) at $T=0$ (dashed) and $T=2\mu/k_B$ (solid). Thin lines are from Eq.~\eqref{e:flucGauss0} at $T=0$ and Eq.~\eqref{e:flucGaussT} at $T=2\mu/k_B$. The vertical black line is where $\min\{L_{x,y}\}=\xi$. Parameters A and B are detailed in Fig.~\ref{f:flucvarlambda}. See text for definitions of $A_\sigma$ and $\xi$. Insets:\,schematic of cells  in the $xy$ plane.}
 \label{f:SchemPhase} 
   \vspace*{-0.25cm}
\end{figure}

Focusing on the  quasi-2D  case of a dipolar BEC tightly confined in one direction we consider the measurement of number fluctuations in cells  (c.f.~fluctuation measurements of a nondipolar BEC in \cite{Hung2011,Hung2013}). We demonstrate that the fluctuations  are anisotropic, as schematically illustrated in Fig.~\ref{f:SchemPhase}(a): Two cells of identical geometry, but orthogonal orientations, measure identical mean atom numbers, however the number variances in these two measurement cells are  strikingly different [Fig.~\ref{f:SchemPhase}(b)].  This effect most strongly manifests itself in the thermally activated collective modes of the system and thus persists even for cell sizes much larger than the correlation length. We show that with increasing cell size the fluctuations of the dipolar system more slowly approach  the thermodynamic regime than a condensate with contact interactions.

{\bf Formalism:}
 We work in a homogeneous quasi-2D geometry by assuming that there
is a strong one-dimensional  harmonic trap of frequency $\omega_z$ in the
$z$-direction, $U(z) = \frac{1}{2}m\omega^2_zz^2$, where $m$ is the atomic  
mass.  The condensate is of the form $\psi_0(\bx)=\sqrt{n}\chi(z)$, where $n$ is the areal density. We will consider the quasi-2D regime \cite{Pedri2005a,Fischer2006a} where the interaction energy scale is small compared to $\hbar\omega_z$  and $\chi(z)$ can be taken as the ground state harmonic oscillator wave function.  The chemical potential, neglecting the $z$ confinement energy, is given by $\mu = n[g_s+(3\cos^2\alpha-1)g_d]$ for dipoles polarised along $\hat{\mathbf{e}} = \hat{\mathbf{z}}\cos\alpha+\hat{\bx}\sin\alpha$ [see Fig.~\ref{f:SchemPhase}(a)],
where  $\alpha$ is the angle between $\hat{\mathbf{e}} $ and  the $z$ axis \cite{Ticknor2011a}. We have introduced $g_s=2\sqrt{2\pi}\hbar^2a_s/ml_z$ as the quasi-2D contact coupling constant, where $a_s$ is the scattering length and $l_z=\sqrt{\hbar/m\omega_z}$ is the axial harmonic oscillator length.  Considering magnetic dipoles with a permanent moment $\mu_m$, the quasi-2D DDI coupling constant is $g_d=\mu_0\mu_m^2/3\sqrt{2\pi}l_z$.  
The excitations are given by  Bogoliubov theory
\begin{equation}
{E_{\bk } }  =  \sqrt{\efree[\efree+2n\tilde{V}(\bk )]},\label{Eq:Bog} 
\end{equation}
 where $\bk =(k_x,k_y)$, $\efree = \hbar^2k^2/2m$, and $n\tilde{V}(\bk )=\mu-3ng_d G(k l_z/\sqrt{2})[\cos^2\alpha-(k_x/k)^2\sin^2\alpha]$ is the  $k$-space interaction potential between the dipoles. We have introduced the function $G(q)\equiv\sqrt{\pi}qe^{q^2}\mathrm{erfc}(q)$, with  $\mathrm{erfc}$ the complementary error function.
 For $\alpha=0$ (untilted dipoles) we have $\tilde{V}(\bk )=g_s+g_d[2-3G(k l_z/\sqrt{2})]$ and the DDIs are momentum dependent, but isotropic.  For $\alpha\ne0$ the anisotropic term $\propto k^2_x/k^2$ contributes.

The \textit{in situ} imaging of ultra-cold gases can be used to measure the atom number within finite-sized cells  determined by the combined properties of the point spread function of the imaging system and the pixels of the camera used to capture the image \cite{Hung2011a,Jacqmin2011a,Armijo2012a}. The operator for atom number within such a cell is  $\hat{N}_\sigma\equiv\int d^2\brho \,\sigma(\brho)\hat{n}(\brho)$, where $\sigma(\brho)$ is the cell weight function, and $\hat{n}$ is the density operator.  Considering imaging along the $z$ direction we  approximate the weight function as the 2D Gaussian $\sigma(\brho) = 2 e^{-(x/L_x)^2-(y/L_y)^2}$ with $1/e$ lengths $L_x$ and $L_y$  along $x$ and $y$, respectively.  This cell has an effective area of $A_\sigma =  2\pi L_xL_y$ and  an aspect ratio of $\lambda\equiv L_y/L_x$. Having anisotropic cells (i.e.~$\lambda\ne1$) is essential to probe the intrinsic anisotropy that emerges in the dipolar BEC and can be realised by amalgamating the signal from multiple pixels, e.g., as was done in Ref.~\cite{Armijo2012a}. 

The mean atom number per cell is $N_\sigma=nA_\sigma$  and only depends on the cell area (i.e.~independent of $\lambda$). 
The variance in atom number $\Delta N_\sigma^2\equiv  \langle (\hat{N}_\sigma- {N}_\sigma)^2\rangle$ is  given by \cite{Astrakharchik2007a,Klawunn2011}
\begin{align}
\Delta N_\sigma^2&= \int d^2\br \,\tau_\sigma(\br) [n^2 g^{(2)}(\br)-n^2 + n\delta(\br)], \\
&= n\int \frac{d^2\bk }{(2\pi)^2}\taut(\bk ) S(\bk ),\label{DN2B} 
\end{align}
where $g^{(2)}(\br)=n^{-2}\langle:\hspace{-0.7mm}\hat{n}(\brho)\hat{n}(\brho')\hspace{-0.7mm}:\rangle$ is the density-density correlation function with $\br=\brho-\brho'$, $S(\bk )=N^{-1}\langle \delta \hat{n}_{\bk } \delta \hat{n}_{-\bk }  \rangle$ is the static structure factor, $\delta \hat{n}_{\bk }=\mathcal{F}\{  \hat{n}(\brho)-{n}\}$ is the density fluctuation operator, with $\mathcal{F}$ the 2D Fourier transformation operator. The cell geometry function is $\tau_\sigma(\br) = \int d^2\brho \int d^2\brho' \sigma(\brho)\sigma(\brho')\delta(\brho-\brho'-\br)$ with transform \cite{Astrakharchik2007a,Klawunn2011}
\begin{equation}
\taut(\bk )=\mathcal{F}\{\tau_\sigma(\br)\} = A_\sigma^2 e^{-[(k_x L_x)^2 + (k_y L_y)^2]/2},\label{e:taugaussian}
\end{equation} 
for the Gaussian cell considered here.
The static structure factor is given by the Feynman relation
\begin{align}
    S(\bk) &= \frac{\efree}{E_\bk}\left(2\bar{n}_{\bk} + 1\right) , \label{Sq}
\end{align}
where $\bar{n}_{\bk}=1/(e^{\beta E_{\bk}}-1)$ and  $\beta=1/\kt$. This  result is valid for dipolar BECs  well below the condensation temperature \cite{Macia2012a} and for quasicondensates \cite{Deuar2009a,Jacqmin2011a}.

In Fig.~\ref{f:SchemPhase}(b) we consider the fluctuations using Eq.~(\ref{DN2B}) as a function of  cell area $A_\sigma$  for cells with long axis parallel to the $x\:(\|)$ or  $y\:(\perp)$ directions. We show results for a tilted $(\alpha=\pi/3)$ and comparable\footnote{The coupling constants are chosen so that both systems have the same values for $\mu$ and $g_d\cos^2\alpha$ and to be reasonably far from the stability boundary.} untilted case ($\alpha=0$).  The tilted results  exhibit significantly larger fluctuations when the long axis of the measurement cell is parallel to the direction along which the dipoles tilt, as compared to the equivalent orthogonally oriented cell. These results show that this anisotropy is strongest for moderate-area cells, and is enhanced with temperature.

The behavior of  $\Delta N_\sigma^2$ is well understood for  cells with dimensions either much larger or much smaller than  the important system  length scales, i.e.~the healing  length $\xi=\hbar/\sqrt{m\mu}$ and the thermal wavelength:\\
\textit{For small cells}  -- the number fluctuations are dominated by the high-$k$ structure factor, which takes the uncorrelated value  $S(\bk)\approx1$ [see Fig.~\ref{f:Sk}]. We can see this by noting that through the geometry function $\taut(\bk )$ the dominant contribution of $S(\bk )$ to the number fluctuations \eqref{DN2B} occurs for wave vectors  $|k_{x,y}|\lesssim1/L_{x,y}$. Thus for small cells the uncorrelated quantum shot noise regime is obtained and the fluctuations are Poissonian, i.e.~$\Delta N_\sigma^2\to N_\sigma$ [see $A_\sigma\to0$ in Fig.~\ref{f:SchemPhase}(b)]. Because it is difficult to resolve cells smaller than the system length scales with optical imaging, it is challenging to directly probe this regime in experiments. \\
\textit{For large cells} -- the fluctuations are determined by the $k\to 0$  structure factor, with  $S(\mathbf{0})=\kt/\mu$ [see Fig.~\ref{f:Sk}]. We thus recover the classical thermodynamic result   $\Delta N_\sigma^2/N_\sigma \to S(\mathbf0)= n\kappa_T \kt$, where $\kappa_T=1/n\mu$ is the isothermal compressibility [i.e.~$T>0$ results  as  $A_\sigma\to\infty$ in Fig.~\ref{f:SchemPhase}(b)].

\begin{figure} 
   \centering
   \includegraphics[width=3.5in]{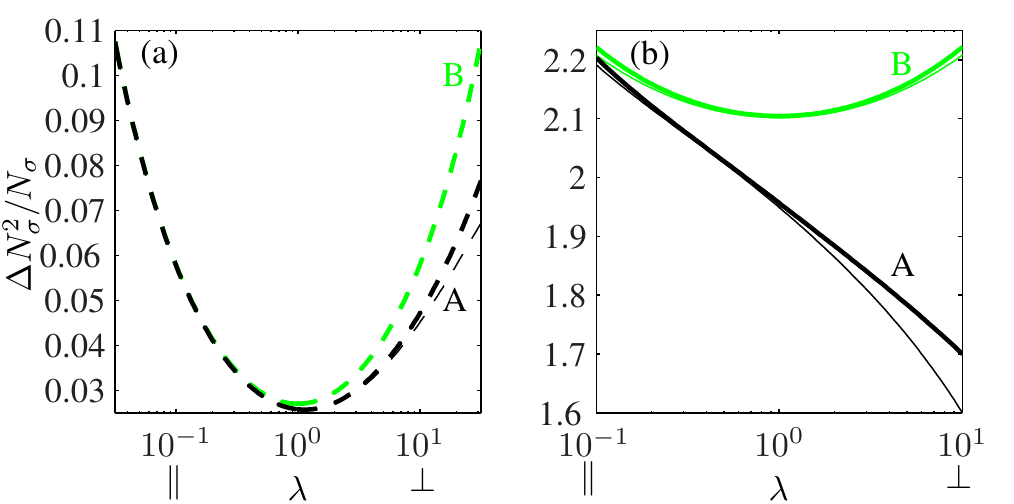} 
   \caption{(color online) Fluctuations as a function of cell aspect ratio with $A_\sigma=(60\xi)^2$ for  parameter set 
   A ($\alpha=\pi/3$, $ng_s=0.25\hbar\omega_z$, $ng_d=0.4\hbar\omega_z$, black) and  
   B ($\alpha=0$,   $ng_s=-0.05\hbar\omega_z$, $ng_d=0.1\hbar\omega_z$, green)  
   at (a) $T=0$ with thin lines  from Eq.~\eqref{e:flucGauss0}, (b) $T=2\mu/k_B$ with thin lines from Eq.~\eqref{e:flucGaussT}.}
   \label{f:flucvarlambda}
\end{figure}

\begin{figure} 
   \centering
  \includegraphics[width=3.4in]{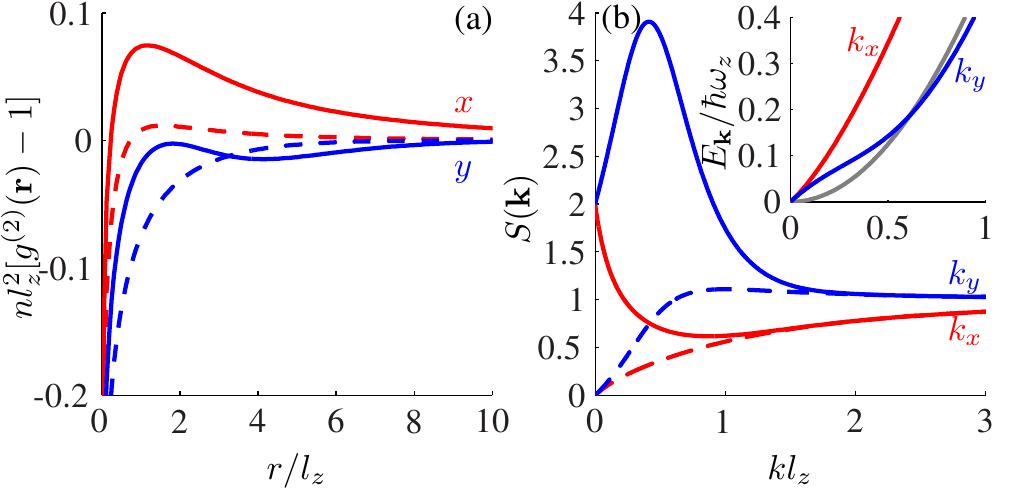}
  \caption{(color online) (a) Density-density correlation function and (b) static structure factors, at $T=0$ (dashed lines) and $T=2\mu/k_B$ (solid lines) for parameter set A defined in Fig.~\ref{f:flucvarlambda}. (inset) Bogoliubov dispersion relation and  free particle dispersion $\efree$ (gray). Along the $x,k_x$-axis (red) and along  the $y,k_y$-axis (blue). Parameter set B results are isotropic and the same as parameter set A along the $k_y$ axis.\label{f:Sk}}
\end{figure}

The small and large cell limits discussed above have no geometry or anisotropy dependence,  features that emerge for moderate-sized  cells.  In Fig.~\ref{f:flucvarlambda} we compute the fluctuation for moderately sized cells of fixed area $A_\sigma=(60\xi)^2$ but varying aspect ratio $\lambda$.
For both tilted and untilted dipoles  the  fluctuations depend on $\lambda$, however we distinguish two effects: \\
\textit{1.~Geometry dependence:} In general the fluctuations in microscopic cells depend on the shape of the cell, in contrast to the classical thermodynamic expression for macroscopic cells in which the fluctuations  only depend on the cell area. Geometry dependence, e.g.~surface area scaling of fluctuations for ideal fermions has been studied previously \cite{Klawunn2011}.\\
\textit{2.~Orientation dependence:} A key prediction is that the tilted dipolar system has anisotropic fluctuations, i.e.~they depend on the cell orientation in the $xy$-plane. In our results [Fig.~\ref{f:flucvarlambda}] this is  revealed by asymmetry of the fluctuations under exchange of the cell dimensions $L_x\leftrightarrow L_y$  (i.e.~$\lambda\leftrightarrow1/\lambda$), particularly obvious at nonzero temperature [Fig.~\ref{f:flucvarlambda}(b)].

We focus on the experimentally relevant regime of cells with dimensions   $L_{x,y} \gg\xi$, where phonon modes dominate, and the role of temperature is characterised by the phonon thermal wavelength $\lambda_T\equiv\hbar^2/mk_BT\xi$ \cite{Armijo2012a,Klawunn2011}. 
We develop an analytic model for this regime using expansions of the structure factors about $k=0$ for $T=0$
\begin{align}
    S(\bk) &=     \frac12 \xi k[1-n \Vdir k/2\mu] + \od{k^3}  , \label{e:Sexpand0}
\end{align}
and for $T>0$
\begin{align}
    S(\bk) &= \frac{\kt}{\mu} [1-n \Vdir k /\mu] + \od{k^2} , \label{e:SexpandT}
\end{align}
where $\Vdir =-3g_dl_z\sqrt{\frac{\pi}{2}}[\cos^2\alpha-(k_x/k)^2\sin^2\alpha]$ is the directional derivative of $\tilde{V}(\bk )$ along  $\bk $ evaluated at $\bk=\mathbf0$. Both results are valid for $k \ll 1/\xi$, and additionally the finite temperature result also requires  $k\ll 1/\lambda_T$. Result (\ref{e:Sexpand0}) is applicable to the  $T\ne0$ regime  for wave vectors $k\gg 1/\lambda_T$, i.e.~for $k$ values of phonons that  are thermally unoccupied. To first order in $\bk$  the quantum fluctuations (\ref{e:Sexpand0}) are isotropic (i.e.~depend only on $k$), whereas the $T>0$ result  (\ref{e:SexpandT}) is anisotropic if $\alpha\ne0$. 
  This anisotropy is apparent in the  $k\to0$ behaviour of $S(\bk)$ in Fig.~\ref{f:Sk}(b): the slope has opposite signs along the $k_x$ and $k_y$ axes for our parameter choice, c.f.~$S(\bk)$ at $T=0$, which is clearly isotropic for small $k$.  
  For the purely contact system ($g_d=0$), $\Vdir=0$ and the leading order correction to $S(\mathbf0)$ for  $T>0$ comes at quadratic order in $k$.

 Evaluating    \eqref{DN2B} using the $T=0$ result \eqref{e:Sexpand0} yields the quantum fluctuations
\begin{align}
\frac{\Delta N^2_\sigma}{N_\sigma}
&\approx   \frac{\xi}{\sqrt{A_\sigma}} \frac{E(1-\lambda^2)}{\sqrt{\lambda}} \notag\\
&\!+\! \frac{ng_d}{\mu} \frac{3(\pi/2)^{3/2}}{A_\sigma /(l_z \xi) } \!\left[\cos^2\alpha \left(\frac{1}{\lambda}\!+\!\lambda\right)-\lambda\sin^2\alpha\right], 
 \label{e:flucGauss0}
\end{align}
valid for $\xi\ll L_{x,y} \ll\lambda_T$, where  $E(u)$ is the complete elliptic integral of the second kind. In this regime, the fluctuations are dominated by   quantum (unoccupied) phonon modes, and to leading order this gives a nonextensive scaling $\Delta N_\sigma^2\propto\sqrt{N}_\sigma$ \cite{Astrakharchik2007a,Klawunn2011}. While accessing this regime is challenging, recent experiments in quasi-1D gases have come close enough to infer quantum corrections \cite{Armijo2012a}. 

In the higher temperature regime we use  result \eqref{e:SexpandT} to find
\begin{align}
  \frac{\Delta N^2_\sigma}{N_\sigma}
  &\approx \frac{\kt}{\mu}\left\{1 + \frac{ng_d}{\mu} \frac{3\sqrt{2\pi}}{\sqrt{ A_\sigma}/l_z}  \left[ \vphantom{\frac{1}{u_x}} \cos^2\alpha \frac{E(1-\lambda^2)}{\sqrt{\lambda}} \right.\right.\notag\\&\left.\left.
  + \sin^2\alpha \frac{  E(1-\lambda^2)-K(1-\lambda^2) }{1-\lambda^2}\lambda^{3/2}  \right]   \right\}, \label{e:flucGaussT}
\end{align}
valid for  $L_{x,y}\gg\max\{\xi,\lambda_T\}$,
where $K(u)$ is the complete elliptic integral  of the first  kind. 
The first right hand term ($\kt/\mu$) is the thermodynamic result for fluctuations, where the rest are leading order finite cell size corrections.

\begin{figure}[!thbp] 
   \centering
   \vspace*{-0.25cm}
    \includegraphics[width=3.4in]{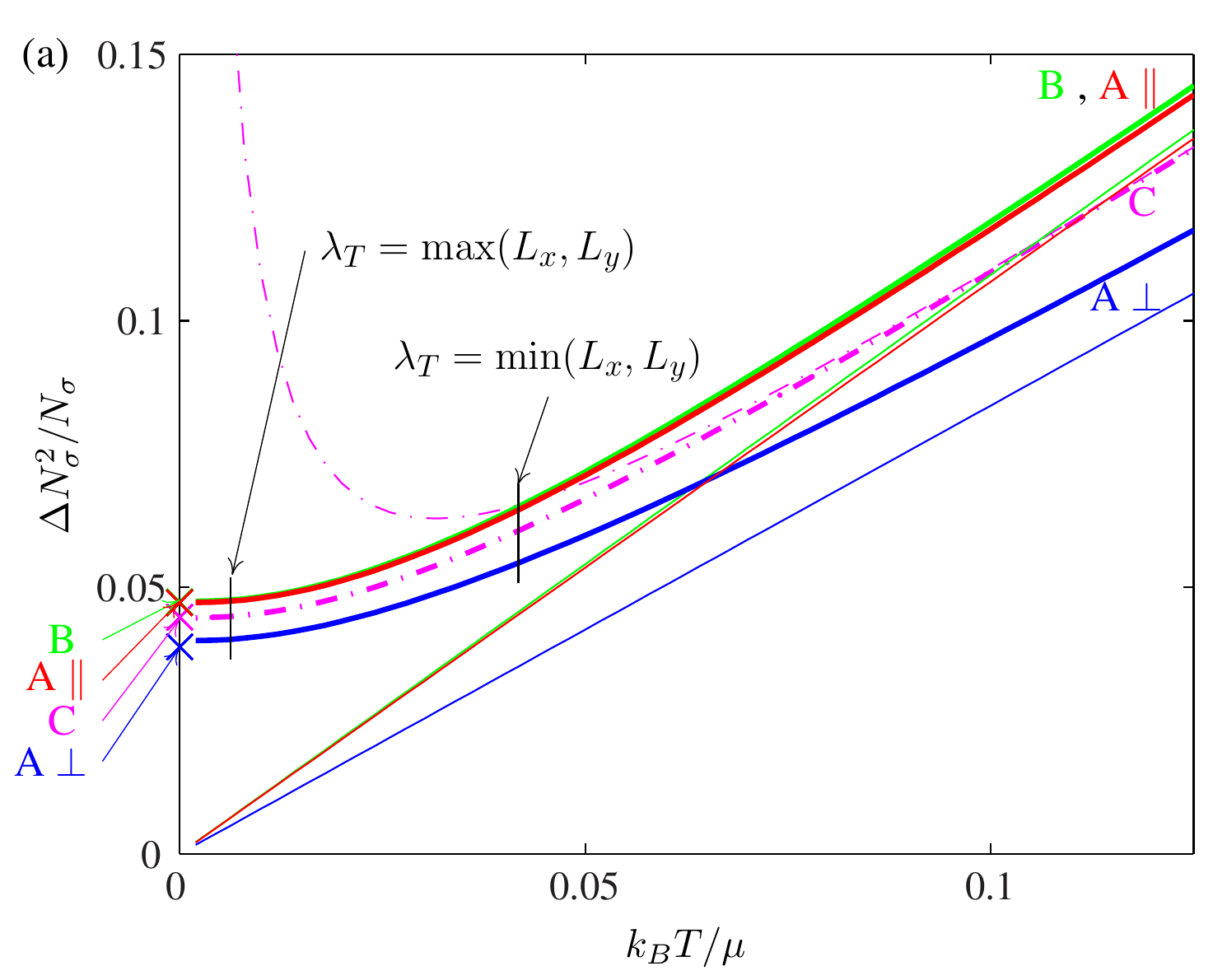} 
    \includegraphics[width=3.4in]{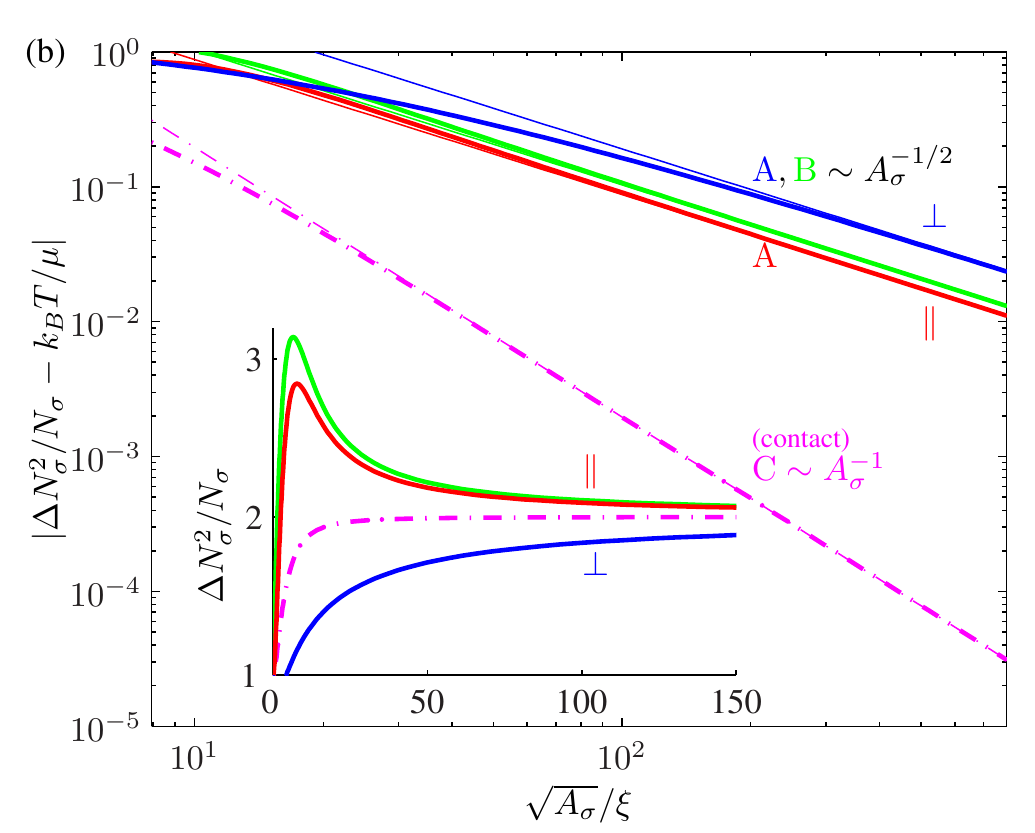} 
   \vspace*{-0.25cm}
   \caption{(color online) (a) Temperature dependence with $A_\sigma=(60\xi)^2$ and (b) large $A_\sigma$ scaling at $T=2\mu/k_B$, of fluctuations as a function of cell area for dipolar BECs with $\alpha=\pi/3$ (A), $\lambda=0.15$ (red) and $\lambda=1/0.15$ (blue) and $\alpha=0$ (B, green), comparing to a contact gas, i.e. $g_d=0$, with same $\mu$ (C, magenta dashed-dotted). Thin lines are the analytic approximation \eqref{e:flucGaussT} for A and B and \eqref{e:flucGaussTcontact} for C. $\times$ shows the analytic approximation \eqref{e:flucGauss0} at $T=0$. Vertical black lines indicate the $T$ values where $\lambda_T=L_x$ and $L_y$, and  mark the transition from quantum to thermal regimes. 
   Parameters A and B are detailed in Fig.~\ref{f:flucvarlambda}.}
   \label{f:flucLscale}
\end{figure}

The predictions of results (\ref{e:flucGauss0}) and (\ref{e:flucGaussT})  are shown in Figs.~\ref{f:SchemPhase} and \ref{f:flucvarlambda}, and are observed to provide a good description of the full numerical results for sufficiently large cells. 
These results also allow us to make some observations about the orientation dependence  (the effect on fluctuations of $\lambda\leftrightarrow1/\lambda$, i.e.~rotating cell by $90^\circ$), noting that $E(1-\lambda^2)/\sqrt{\lambda}=E(1-\lambda^{-2})\sqrt{\lambda}$:
(i)  There is no orientation dependence for a contact gas ($g_d=0$) or for an untilted dipolar gas ($\alpha=0$) and by symmetry this is true for any cell size; (ii) The orientation dependence is enhanced for small $\mu$, however we note that we require $\mu\ge0$ for the system to be stable.

The variation of fluctuations with temperature is shown in Fig.~\ref{f:flucLscale}(a) revealing the transition from the $T=0$ limits given by \eqref{e:flucGauss0} to the near linear dependence predicted by (\ref{e:flucGaussT}) at higher temperature.

Importantly our results demonstrate a key aspect of   DDIs is to  slow the approach to the thermodynamic result with cell size. This is demonstrated in Fig.~\ref{f:flucLscale}(b) by comparing dipolar condensates to a condensate  with purely contact interactions. Whereas the finite size corrections to the dipolar system scale as $\sim\!A_\sigma^{-1/2}$ as clearly seen from \eqref{e:flucGaussT}, the contact case converges more rapidly as 
$\sim\! A_\sigma^{-1}$. Indeed, for the contact case the fluctuations including the leading order finite size correction are
\begin{align}
    \frac{\Delta N^2_\sigma}{N_\sigma} &\approx \frac{\kt}{\mu} +  \frac{\pi\xi^2}{2A_\sigma}\left(\frac{\mu}{3k_BT}-\frac{k_BT}{\mu}\right)\left(\lambda+\frac{1}{\lambda}\right) \label{e:flucGaussTcontact}.  
\end{align}

{\bf Conclusions and outlook:}
We have developed a  theory of fluctuation measurements made in dipolar condensates with finite-sized anisotropic measurement cells, providing analytic results for quantum and thermal fluctuations.
A key prediction is that, by tilting the dipoles, anisotropy in the fluctuations arise. Our results show that this should be easily accessible in experiments, because (i) the system does not need to be close to the stability boundary (as is required to see rotons \cite{Santos2003a,Bisset2013b}) and (ii) the effect persists for large cells (much greater than healing length), and thus does not require high numerical aperture imaging. We have shown that this latter effect is assisted by the slow onset of the thermodynamic limit (where anisotropy of fluctuations vanishes) as cell size increases, arising from the long range DDI. Through the use of a local density approximation  \cite{Jacqmin2011a,Armijo2012a} (validated for DDIs in Refs.~\cite{Blakie2012a,Bisset2013a}) our result is applicable to condensates in pancake shaped traps, or as in Ref.~\cite{Hung2011,Hung2013} to the central region of such a system where the average density is approximately constant. Our predictions will be realisable with current experiments, e.g.~parameter set A corresponds to  $N=3.5\times10^4$ $^{164}$Dy atoms with $a_s=4$\:nm in a  $(\omega_{\rho},\omega_z)\!=\!2\pi\!\times\!(10,1900)\,$s$^{-1}$ trap (c.f.~\cite{Hung2011}), and the results in Fig.~\ref{f:flucvarlambda} are for a cell of area  $A_\sigma\approx 800\, \mu\mathrm{m}^2$.
The temperatures we have considered are less than $0.4T_c^0$, with $T_c^0=\sqrt{6N}\hbar\omega_{\rho}/\pi k_B$ the ideal condensation temperature. We have also validated the slow approach to the thermodynamic limit in the central region of the trapped system (where the mean density does not vary significantly) with full calculations including the trapping potential.

\noindent {\bf Acknowledgments:}
DB and PBB acknowledge support by the Marsden Fund of New Zealand (contract UOO1220).
RNB and CT acknowledge support from CNLS, LDRD, and LANL which is operated by LANS, LLC for the NNSA of the US DOE (contract no.~DE-AC52-06NA25396).

\bibliographystyle{apsrev4-1}

\end{document}